# An Adjustable Chance-Constrained Approach for Flexible Ramping Capacity Allocation

Zhiwen Wang, *Student Member, IEEE*, Chen Shen, *Senior Member, IEEE*, Feng Liu, *Member, IEEE,* Jianhui Wang, *Senior Member, IEEE*, Xiangyu Wu, *Student Member, IEEE*

**Abstract**—With the fast growth of wind power penetration, power systems need additional flexibility to cope with wind power ramping. Several electricity markets have established requirements for flexible ramping capacity (FRC) reserves. This paper addresses two crucial issues that have rarely been discussed in the literature: 1) how to characterize wind power ramping under different forecast values and 2) how to achieve a reasonable trade-off between operational risks and FRC costs. Regarding the first issue, this paper proposes a concept of conditional distributions of wind power ramping, which is empirically verified by using simulation and real-world data. For the second issue, this paper develops an adjustable chance-constrained approach to optimally allocate FRC reserves. Equivalent tractable forms of the original problem are devised to improve computational efficiency. Tests carried out on a modified IEEE 118-bus system demonstrate the effectiveness and efficiency of the proposed method.

## NOMENCLATURE

Most symbols used in this paper are listed below; others are defined following their first appearance.

### Indices and numbers

| | |
|---|---|
| $i, g, w, d,$ $m, l$ | Indices for periods, conventional units, wind farms, loads, Gaussian components, and transmission lines |
| $I, G, W, D,$ $M, L$ | Number of periods, conventional units, wind farms, loads, Gaussian components, and transmission lines |

### Parameters

| | |
|---|---|
| $p_{w,i}/p_{w,i}^{\text{fore}}$, $p_i^W$ | Actual/forecast power of wind farm $w$, and the aggregated actual wind power |
| $p_{d,i}$, $p_i^D$ | Power of load $d$, and the aggregated load power |
| $ND_i$, $\Delta ND_i^{i+1}$ | Net load and its ramping |
| $p_g^{\max}$, $p_g^{\min}$ | Limits of unit $g$ |
| $r_{g,i}^{\text{up\_lim}}$, $r_{g,i}^{\text{dn\_lim}}$ | Limits of up/down flexible ramping capacity (FRC) reserves |
| $F_l^{\lim}$ | Limits of line power |
| $\lambda$ | Cost coefficients of FRC reserves, and penalty coefficients of wind spillage/load shedding |
| $\omega_m, \boldsymbol{\mu}_m, \boldsymbol{\Sigma}_m$ | Weight coefficient, mean vector, and correlation matrix of a Gaussian component |
| $a_g, b_g, c_g$ | Cost coefficients of unit $g$ |
| $s_{l,g}, s_{l,w}, s_{l,d}$ | Generation shift factors of line $l$ |
| $\boldsymbol{X}, \boldsymbol{Y}$ | Wind power and corresponding forecasts |
| $\Delta \boldsymbol{X}, \Delta \boldsymbol{Y}$ | Wind power ramping and corresponding forecasts |
| $\boldsymbol{Z}$ | Random variables in a compact form |

### Functions

| | |
|---|---|
| $N_m(\cdot)$ | Multivariate normal distribution |
| $\det(\cdot)$ | Determinant of a matrix |
| $\text{pdf}(\cdot)$ | Probability density function (PDF) |
| $\text{cdf}(\cdot)$ | Cumulative distribution function (CDF) |
| $F_l(\cdot)$ | Linear map function transforming power injections into line power |

### Decision Variables

| | |
|---|---|
| $p_{g,i}$ | Scheduled power of unit $g$ |
| $r_{g,i}^{\text{up}}, r_{g,i}^{\text{dn}}$ | Scheduled up/down FRC reserves |
| $\alpha, \beta$ | Confidence level and its quantile |
| $\boldsymbol{u}$ | Decision variables in a compact form |

## I. INTRODUCTION

INTEGRATING a large amount of wind power into power systems is a crucial, albeit challenging, issue. Variations of wind power within a short period, known as "wind power ramping" [1], may exhaust reserves, consequently cause undesirable load shedding and/or wind spillage, and increase operational risks/costs of power systems. In this context, much effort has been devoted to two key topics: characterizing wind power ramping and developing appropriate methods that can

This work was supported in part by the Foundation for Innovative Research Groups of the National Natural Science Foundation of China under Grant 51621065, and in part by the Special Fund of the National Basic Research Program (973) of China under Grant 2013CB228201. (*Corresponding to: Chen Shen*)

Z. Wang, C. Shen, and F. Liu are with the State Key Lab. of Power Systems, Department of Electrical Engineering, Tsinghua University, Beijing 100084, China (e-mails: wang-zw13@mails.tsinghua.edu.cn, shenchen@mail.tsinghua.edu.cn, lfeng@mail.tsignhua.edu.cn).

J. Wang is with the Department of Electrical Engineering at Southern Methodist University, Dallas, TX, USA and the Energy Systems Division at Argonne National Laboratory, Argonne, IL, USA (jianhui.wang@anl.gov).

X. Wu is with the School of Electrical Engineering, Beijing Jiaotong University (BJTU), Beijing 100044, China (e-mail: wuxiangyu639@163.com).

mitigate the detrimental effects of ramping.

### A. Literature review

In order to model wind power ramping, an optimal detection technique has been proposed to identify ramping events from data series [2]. Aiming at a similar purpose, swinging-door algorithms [3-5] have been developed for ramping-event detection and prediction. Further, a neural-network-based method has been proposed for forecasting and generating ramping scenarios [6]. A comprehensive review on this topic can be found in [7]. When a number of ramping events are detected from historical data or simulated by numerical approaches, distributions of wind power ramping can be obtained by statistics. Interestingly, according to the results in [2], [6], [8], and [9], wind power ramping does not follow a Gaussian distribution. Although significant progress has been made in modeling wind power ramping [2-9], the following two questions need to be further investigated:

(1) Given an effective forecasting tool, different forecast ramping values can imply different chances of ramping events. For example, a small forecast ramping value indicates a good chance of a small-magnitude ramping event, and vice versa. This fact motivates researchers to characterize wind power ramping under different forecast values. In this regard, the first question is how to model conditional distributions of wind power ramping with respect to different forecast values?

(2) To incorporate conditional distributions of wind power ramping into unit commitment (UC) and economic dispatch (ED), the second question is how to choose an appropriate form of the conditional distribution that can facilitate the decision-making for UC and ED?

Currently, with the goal of coping with wind power ramping, the FRC allocation has been considered in generation schedules and/or the market clearing process. In CAISO, FRC products are launched in the real-time (RT) market [10]. In MISO, FRC products are procured for both the RT and day-ahead markets [11]. In industrial practice and in the literature, there are several commonly used approaches to allocate FRC:

(1) *Fixed FRC requirements*. In CAISO, the FRC reserves consist of two parts: a portion due to net load forecast change and a portion due to ramping uncertainty within a confidence interval from 2.5% to 97.5% [9]. With this approach, 95% ramping scenarios are expected to be handled. In MISO, following the *Gaussian-sigma rule*, FRC reserves are scheduled to cover 2.5 standard deviations (99% confidence levels) of forecasts [11]. Similar approaches are used in [12] and [13]. A drawback of these approaches is that fixed requirements are either so strict that FRC reserves are overcommitted with concomitant high costs, or so lax that undercommitted FRC reserves may not cope with possible wind power ramping events.

(2) *Scenario-based stochastic optimization*. In [14], a stochastic real-time unit commitment method is proposed to evaluate FRC market design. In a relevant study [15] based on the scenario optimization method, energy storage systems are used to limit ramp rates of wind power. These scenario-based methods can convert the stochastic optimization problem into a deterministic one. However, a major issue is that the deterministic programming model could turn out to be intractable as the sample size becomes very large.

(3) *Robust optimization*. Compared with the scenario-based approaches, robust optimization has an advantage in computational tractability. In [16], with consideration of severe power-ramping events in daily operation, a robust model for multi-year planning was developed. In [17], a robust optimization framework was established to address the deliverability issue of FRC. To limit operational risks brought about by volatile renewables, a robust risk-constrained UC formulation was proposed in [18]. Since robust optimization is usually focused on the worst-case scenario, it may cause high costs, i.e., it is conservative.

(4) *Chance-constrained programming* (CCP). CCP provides a promising alternative to deal with uncertainties [19-29] by allowing violation of constraints within a tolerable probability, say, 5%. In [19], a tractable chance-constrained ED is presented. First, the uncertainties of multiple wind farms are assumed to be Gaussian and independent from each other. Then, the chance-constrained transmission line limits are converted into a set of second-order cone inequalities. As a result, the original problem is converted into a second-order cone programming problem, which can be solved efficiently. A drawback of [19] is that the Gaussian assumption of uncertainties may cause inaccuracy in the modeling and the optimal solutions. In related studies [20-22], stochastic loads and wind generation are modeled by Gaussian distributions. Extending the work of [19], Lubin et al. [23] developed a robust chance-constrained method, where the mean and variance of a Gaussian distribution are not fixed, but lie within a given uncertainty set. As wind power ramping is inherently non-Gaussian, the Gaussian-based methods [19-23] may not directly apply.

The CCP methods with non-Gaussian models are studied in [24-29]. In [24], a non-parametric approach is proposed. First, random variables, e.g., solar power, are modeled by discrete empirical distributions. Then, chance constraints are computed using the discrete convolution technique. In [25], [26], and [27], chance constraints are approximated by using a number of scenarios. Since these approaches do not have analytical formulae, they appear to be less efficient than the parametric methods. In [28], a two-parameter Weibull distribution was used to model wind speed. Then, the distribution of wind power was obtained as a variant of the Weibull distribution. The sufficient-generation requirement was formulated as a chance constraint, and was converted into an equivalent linear inequality. In [29], the so-called "Versatile distribution" was proposed to characterize the nature of wind power uncertainty. The regulation reserve limits were formulated as chance constraints. Compared with the Weibull distribution, the Versatile distribution admits analytical CDF and inverse CDF. Hence, quantiles of chance constraints and derivatives of the objective function can be easily obtained, and are expected to significantly improve the computational efficiency when solving chance-constrained ED problems.

Although significant progress has been described in the literature [19-29], a common drawback of these CCP methods is that the tolerable violation probabilities, i.e., confidence levels, of chance constraints rely on personal experience. Usually, they are predefined values, e.g., 5%. Intuitively, a small value for confidence level can restrain the operational risk, such as wind spillage and load shedding, but increase the cost, and vice versa. Unfortunately, how to appropriately select confidence levels for chance constraints remains an open question.

### B. Contributions

As mentioned previously, there is little research that addresses the modeling of conditional wind power ramping and the balancing of risks and FRC costs in generation scheduling. To close this gap, this paper studies a risk-adjustable FRC allocation approach. Its main contributions are threefold:

(1) In terms of problem formulation, a risk-adjustable chance-constrained optimization model for FRC allocation is proposed. The model is able to find optimal confidence levels, achieving a reasonable trade-off between FRC reserve costs and potential losses due to wind spillage/load shedding.

(2) In uncertainty modeling, a conditional-distribution-based model of wind power ramping is proposed to characterize ramping under different forecast ramping values.

(3) For the solution methodology, adjustable chance constraints are converted into a set of linear inequalities. Analytical formulae for integral terms of the objective function are derived. As a consequence, the original problem has a tractable form. The computational efficiency is thereby improved.

### C. Organization

The rest of this paper is organized as follows: Section II presents the problem formulation. In Section III, conditional distributions of wind power ramping are detailed. In Section IV, a solution methodology for the original problem is discussed. In Section V, the test results are presented. Section VI draws conclusions, with a discussion of limitations.

## II. PROBLEM FORMULATION

This section provides the formulation of the FRC allocation problem. First, three concepts are introduced. They are the net load ramping, adjustable chance constraints, and potential losses. Then, the objective function is defined, followed by a description of constraints. Possible extensions and a general compact form for the optimization model are discussed. Finally, challenges of the problem are pointed out.

### A. Net load ramping

Net load is defined as the total load demand minus the aggregated wind power:

$$ND_i = p_i^D - p_i^W \qquad (1)$$

$$p_i^D = \sum_{d=1}^{D} p_{d,i} \qquad (2)$$

$$p_i^W = \sum_{w=1}^{W} p_{w,i} \qquad (3)$$

There are different definitions of the net load ramping [1].

Since this paper is concerned with the reserve allocation, we adopt a definition that is used in ED by CAISO [9] and MISO [10]. Net load ramping is defined as follows:

$$\Delta ND_i^{i+1} = ND_{i+1} - ND_i \qquad i = 1, \cdots, I\text{-}1 \qquad (4)$$

### B. Adjustable chance constraints

Sufficient FRC up/down reserves should be allocated to cover possible wind power ramping. As the net load ramping is random, FRC requirements are formulated as chance constraints:

$$\Pr\left( \Delta ND_i^{i+1} \leq \sum_{g=1}^{G} r_{g,i}^{up} \right) \geq 1 - \alpha_i^{up} \quad \forall i \qquad (5)$$

$$\Pr\left( -\sum_{g=1}^{G} r_{g,i}^{dn} \leq \Delta ND_i^{i+1} \right) \geq 1 - \alpha_i^{dn} \quad \forall i \qquad (6)$$

Equations (5) and (6) indicate that there is a $(1-\alpha)$ chance that the net load ramping can be covered by the scheduled FRC reserves. In the literature [19-27], the confidence level $\alpha$ is a small fixed number, e.g., 5%. However, fixed confidence levels suffer from several problems:

(1) A small $\alpha$ may lead to a very high control cost. Worse still, a small $\alpha$ may result in infeasibility of the problem.

(2) Determining the value of $\alpha$ depends on personal experience. A lower $\alpha$ leads to a more secure system, while resulting in a higher economic cost. In fact, operators don't know what value of $\alpha$ is optimal regarding either security or economy. There is no standard to follow.

To solve these problems, the concept of "adjustable chance constraint" is proposed, which is defined as follows:

An adjustable chance constraint means the confidence level $\alpha$ is not a predefined parameter, but a decision variable. An optimization problem with adjustable chance constraints is to find optimal confidence levels $\alpha$ and other decision variables with which the objective function is minimal.

To optimally determine the confidence levels, the costs caused by violations of chance constraints should be quantified. To this end, a concept of potential losses is introduced below.

### C. Potential losses

A potential loss refers to the expected penalty for violations of chance constraints. Take the chance constraint of Eq. (5) as an example. The PDF of the net load ramping is shown in Fig. 1. The chance constraint (Eq. (5)) indicates that there is $\alpha$ chance that the ramping requirement is not satisfied. If up ramping reserves are not enough, the increasing net load demand cannot be fully satisfied. As a result, load shedding is activated. In this case, the potential loss is defined as the expected penalty for load shedding.

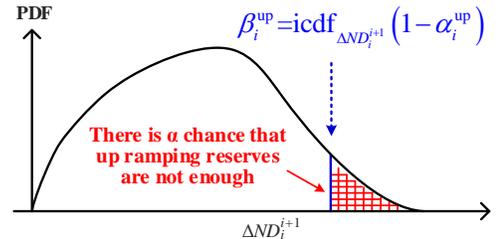

Fig. 1 Illustration of the potential loss due to insufficient up ramping reserves.

Denote the $(1-\alpha_i^{up})$ quantile of $\Delta ND_i^{i+1}$ by $\beta_i^{up}$:

$$\beta_i^{up} = \mathrm{icdf}_{\Delta ND_i^{i+1}}\left(1-\alpha_i^{up}\right) \qquad (7)$$

The potential loss is computed as follows:

$$\mathbf{E}\left[f_{Ramp}^{up}\right] = \sum_{i=1}^{I}\lambda_{shed}\int_{\beta_i^{up}}^{max}\left(\nu-\beta_i^{up}\right)\mathrm{pdf}_{\Delta ND_i^{i+1}}\left(\nu\right)d\nu \qquad (8)$$

where $\mathrm{pdf}_{\Delta ND_i^{i+1}}(\cdot)$ is the PDF of $\Delta ND_{i+1,i}$; $\mathrm{icdf}_{\Delta ND_i^{i+1}}(\cdot)$ is the inverse CDF of $\Delta ND_i^{i+1}$.

Similar results can be obtained for the chance constraint of Eq. (6). If down ramping reserves are not sufficient, the decrease in the net load leads to excess power. Hence, a portion of the wind power should be curtailed. In this case, the potential loss is defined as the expected penalty for wind spillage, i.e.,

$$\mathbf{E}\left[f_{Ramp}^{dn}\right] = \sum_{i=1}^{I}\lambda_{spill}\int_{\beta_i^{dn}}^{max}\left(\nu-\beta_i^{dn}\right)\mathrm{pdf}_{-\Delta ND_i^{i+1}}\left(\nu\right)d\nu \qquad (9)$$

$$\beta_i^{dn} = \mathrm{icdf}_{-\Delta ND_i^{i+1}}\left(1-\alpha_i^{dn}\right) \qquad (10)$$

where $\mathrm{pdf}_{-\Delta ND_i^{i+1}}(\cdot)$ is the PDF of $-\Delta ND_{i+1,i}$; $\mathrm{icdf}_{-\Delta ND_i^{i+1}}(\cdot)$ is the inverse CDF of $-\Delta ND_i^{i+1}$.

### D. Objective function

The objective function consists of two terms: the potential losses and deterministic control costs.

$$f_{total} = \sum_{i=1}^{I}\sum_{g=1}^{G}\left(C_{g,i}^{F}+C_{g,i}^{R}\right)+\mathbf{E}\left[f_{Ramp}^{dn}+f_{Ramp}^{up}\right] \qquad (11)$$

$$C_{g,i}^{F} = a_g\,p_{g,i}^2+b_g\,p_{g,i}+c_g \qquad \forall g,\forall i \qquad (12)$$

$$C_{g,i}^{R} = \lambda_g^{UP}\,r_{g,i}^{up}+\lambda_g^{DN}\,r_{g,i}^{dn} \qquad \forall g,\forall i \qquad (13)$$

The optimization problem aims to find optimal control actions ($r_{g,i}^{up}$, $r_{g,i}^{dn}$, $p_{g,i}$) and confidence levels $\alpha$ with which the sum of the potential losses and deterministic control costs is minimal. When an optimal solution is found, there is no incentive to increase or decrease the value of $\alpha$ because either action will deteriorate the objective function. At this point, $\alpha$ is the optimal confidence level. In this regard, a reasonable trade-off between risks and economy is obtained.

### E. Deterministic constraints

To focus the FRC issue, other constraints are formulated as deterministic ones. They are listed as follows:

1) *Power balance equation*

$$\sum_{g=1}^{G}p_{g,i}+\sum_{w=1}^{W}p_{w,i}^{fore}=\sum_{l=1}^{D}p_{d,i} \quad \forall i \qquad (14)$$

2) *Transmission limits*

$$-F_l^{lim}\le F_l\left(p_{g,i}\,,p_{w,i}^{fore}\,,p_{d,i}\right)\le F_l^{lim} \quad \forall i,\forall l \qquad (15)$$

3) *Ramping capacity limits*

$$0\le r_{g,i}^{up}\le\min(r_{g,i}^{up\_lim},p_g^{max}-p_{g,i}) \qquad \forall g,\forall i \qquad (16)$$

$$0\le r_{g,i}^{dn}\le\min(r_{g,i}^{dn\_lim},p_{g,i}-p_g^{min}) \qquad \forall g,\forall i \qquad (17)$$

4) *Power output limits*

$$p_g^{min}\le p_{g,i}\le p_g^{max} \qquad \forall g,\forall i \qquad (18)$$

5) *Generation movement between periods*

$$-r_{g,i}^{dn}\le p_{g,i+1}-p_{g,i}\le r_{g,i}^{up} \qquad \forall g,\forall i \qquad (19)$$

In this paper, it is assumed that wind power is random, while the load demand is deterministic. This assumption is justified as follows. Usually, wind power uncertainty could be $25-40\%$ of the installed capacity, which is non-negligible. Compared with wind power forecasting, load forecasting is much more accurate. Therefore, the load demand is assumed to be known and deterministic. In order to focus on the FRC requirements, other reserve requirements, e.g., regulating reserves [30], which can be considered in a similar manner, are not included in the model.

### F. Possible modeling extensions

1) *Different ramping intervals*

According to Eq. (4), ramping refers to a power variation between the period $i$ and the next period $i+1$, i.e., the ramping interval is 1. In CAISO, the ramping interval may assume multiple values, say, 1, 2, and 3 [10]. In this case, the ramping definition (4) is modified as follows:

$$\Delta ND_i^{i+\tau}=ND_{i+\tau}-ND_i \qquad \tau=1,2,3 \qquad (20)$$

If Eq. (20) is adopted, ramping capacity requirements (Eqs. (5) and (6)) should be modified accordingly.

2) *Extension to zonal reserve requirements*

Constraints (5) and (6) only enforce the aggregated net load ramping at the system level. Sometimes, there is more wind power in this zone while there is less in another zone. To address this issue, some ISOs require that FRC reserves be allocated zone by zone. This is called zonal reserve requirement [31], [32].

Denote the net load ramping of the $j$th zone by $\Delta ND_i^{i+1}(j)$:

$$\Delta ND_i^{i+1}\left(j\right)=\left(\sum_{d\in D_j}p_{d,i+1}-\sum_{w\in W_j}p_{w,i+1}\right)$$
$$-\left(\sum_{d\in D_j}p_{d,i}-\sum_{w\in W_j}p_{w,i}\right) \qquad (21)$$

Adjustable chance constraints of the $j$th zonal reserves are formulated as follows:

$$\Pr\left(\Delta ND_i^{i+1}\left(j\right)\le\sum_{g\in G_j}r_{g,i}^{dn}\right)\ge1-\alpha_i^{up} \qquad (22)$$

$$\Pr\left(-\sum_{g\in G_j}r_{g,i}^{up}\le\Delta ND_i^{i+1}\left(j\right)\right)\ge1-\alpha_i^{dn} \qquad (23)$$

3) *Extension to transmission limit chance constraints*

In addition to FRC requirements, some deterministic constraints can be formulated as adjustable chance constraints in a similar manner. For example, when transmission congestion is considered, deterministic line limits (15) can be modified as adjustable chance constraints. A detailed discussion of this issue is provided in Appendix A.

### G. General compact form

For brevity, a general compact form for the problem described in Eqs. (4) through (19) is given as follows:

$$\min_{u,\alpha_k}\quad u^{\mathrm{T}}Qu+Ru+\sum_k\lambda_k\int_{\beta_k}^{max}\left(\nu-\beta_k\right)\mathrm{pdf}_{D_k^{\mathrm{T}}Z}\left(\nu\right)d\nu \quad (24)$$

$$s.t. \qquad\qquad Au=b \qquad (25)$$

$$Bu\le c \qquad (26)$$

$$\Pr\left\{\boldsymbol{D}_k^{\mathrm{T}}\boldsymbol{Z} \leq \boldsymbol{C}_k^{\mathrm{T}}\boldsymbol{u} + d_k\right\} \geq 1 - \alpha_k \quad \forall k \quad (27)$$

$$\beta_k = \mathrm{icdf}_{\boldsymbol{D}_k^{\mathrm{T}}\boldsymbol{Z}}\left(1 - \alpha_k\right) \quad (28)$$

where $\boldsymbol{u}$, $\alpha_k$ are decision variables, i.e., net load ramping; $\boldsymbol{Q}$, $\boldsymbol{R}$, $\boldsymbol{A}$, $\boldsymbol{B}$, $\boldsymbol{C}_k$, $\boldsymbol{D}_k$, $\boldsymbol{b}$, $\boldsymbol{c}$, $d_k$ are coefficient matrix/vector/constant with proper dimensions; $\boldsymbol{D}_k^{\mathrm{T}}\boldsymbol{Z}$ represents a random variable; and $\mathrm{icdf}_{\boldsymbol{D}_k^{\mathrm{T}}\boldsymbol{Z}}$ is the inverse CDF of $\boldsymbol{D}_k^{\mathrm{T}}\boldsymbol{Z}$.

Equation (25) represents the power balance requirement (Eq. (14)). Equation (26) represents the linear inequalities of Eqs. (15)−(19). When the chance-constrained FRC requirements are considered, Eq. (27) represents Eqs. (5) and (6). $\boldsymbol{Z}$ has two entries, $\Delta ND_t^{i+1}$ and $-\Delta ND_t^{i+1}$. $\boldsymbol{D}_k$ is a two-dimensional vector consisting of 1 and 0. When transmission limits are considered, Eq. (27) represents the FRC requirements (Eqs. (5)(6)) and the chance-constrained transmission limits (Eqs. (52)(53)) in Appendix A. $\boldsymbol{Z}$ represents the net load ramping in Eqs. (5) and (6) and wind power in Eqs. (52) and (53). The entries of $\boldsymbol{D}_k$ and $\boldsymbol{C}_k$ associated with net load ramping are 1 and 0, while the entries associated with wind power are generation shift factors.

### H. Challenges

In solving the problem defined by Eqs. (24)−(28), there are three major challenges:

First, an appropriate model of net load/wind power ramping is needed. Here, "appropriate" means that the model can accurately characterize the stochastic nature of ramping, and can be easily incorporated into the problem (Eqs. (24)−(28)).

Second, since the adjustable chance constraints cannot be directly computed by commercial solvers, they must be converted to equivalent tractable forms, e.g., linear inequalities.

Third, during an iterative solution of Eqs. (24)−(28), potential losses, as well as their derivatives, should be computed efficiently.

## III. CONDITIONAL DISTRIBUTION OF WIND POWER RAMPING

### A. Benefits of conditional models

Before details of conditional wind power ramping models are given, an example is provided in Fig. 2 to show the benefits of using conditional models in decision-making.

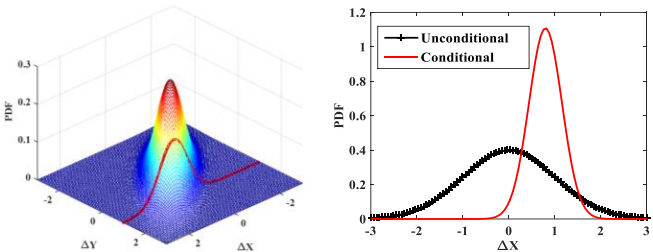

(a) A joint distribution    (b) Unconditional and conditional distributions
Fig. 2 Distributions of $\Delta X$ and $\Delta Y$

Suppose that there are two random variables, $\Delta X$ and $\Delta Y$. They follow a bivariate norm distribution, which is shown in Fig. 2(a). The unconditional distribution of $\Delta X$, regardless of $\Delta Y$, is shown as the black curve in Fig. 2(b). The unconditional distribution is also known as a marginal distribution of $\Delta X$. The black curve indicates that $\Delta X$ ranges from -3.0 to 3.0. Suppose that $\Delta Y$ is known and equal to 1.0. Then, a conditional distribution of $\Delta X$ with respect to $\Delta Y=1$ is obtained, which is shown as the red curve in Fig. 2(a) and Fig. 2(b). The red curve indicates that $X$ ranges from -0.3 to 2.0.

When the unconditional distribution is used in an ED, an operator has to schedule 3.0 MW up reserves and 3.0 MW down reserves to cope with a possible realization of $\Delta X$. As a comparison, when the conditional distribution $\Delta X/\Delta Y=1$ is used, only 2.0 MW down reserves and 0.3 MW up reserves are needed. From this example, it can be seen that the conditional model leads to a lower reserve cost than the unconditional one.

In the following, let $Y$ denote the forecast wind power, $X$ denote the actual wind power, $\Delta Y$ denote the forecast wind power ramping, and $\Delta X$ denote the actual wind power ramping.

Currently, many ISOs have wind power forecasting tools, which can produce forecast values of wind power in advance. That is, $Y$ and $\Delta Y$ are known. From the forecast information of $Y$ and $\Delta Y$, a conditional model of wind power ramping $\Delta X/\Delta Y$, as well as the net load ramping, can be obtained. Details are provided below.

### B. Modeling wind power with Gaussian mixture model

Let a random variable $X_i$ denote the actual wind power output in period $i$, and $Y_i$ denote the corresponding forecast value. Then, two random vectors, $\boldsymbol{X}$ and $\boldsymbol{Y}$, for actual outputs/forecasts over $I$ periods are defined as follows:

$$\boldsymbol{X} = \begin{bmatrix} X_1 & \cdots & X_i & \cdots & X_I \end{bmatrix}^{\mathrm{T}} \quad (29)$$

$$\boldsymbol{Y} = \begin{bmatrix} Y_1 & \cdots & Y_i & \cdots & Y_I \end{bmatrix}^{\mathrm{T}} \quad (30)$$

The Gaussian mixture model (GMM) is adopted to represent a joint distribution of $[\boldsymbol{X}^{\mathrm{T}} \ \boldsymbol{Y}^{\mathrm{T}}]^{\mathrm{T}}$ because the GMM is able to accurately characterize non-Gaussian correlated random variables [33-35]. A GMM is defined as follows:

$$f_{XY}\left(\boldsymbol{x}, \boldsymbol{y}\right) = \sum_{m=1}^{M} \omega_m N_m\left(\boldsymbol{x}, \boldsymbol{y} \ ; \ \boldsymbol{\mu}_m, \boldsymbol{\sigma}_m\right) \quad (31)$$

$$N_m\left(\boldsymbol{x}, \boldsymbol{y} \ ; \ \boldsymbol{\mu}_m, \boldsymbol{\sigma}_m\right) = \frac{e^{-\frac{1}{2}\left(\begin{bmatrix} \boldsymbol{x} \\ \boldsymbol{y} \end{bmatrix} - \boldsymbol{\mu}_m\right)^{\mathrm{T}} \boldsymbol{\sigma}_m^{-1}\left(\begin{bmatrix} \boldsymbol{x} \\ \boldsymbol{y} \end{bmatrix} - \boldsymbol{\mu}_m\right)}}{\left(2\pi\right)^I \det\left(\boldsymbol{\sigma}_m\right)^{1/2}} \quad (32)$$

$$\sum_{m=1}^{M} \omega_m = 1, \quad \omega_m > 0 \quad (33)$$

When historical $[\boldsymbol{X}^{\mathrm{T}} \ \boldsymbol{Y}^{\mathrm{T}}]^{\mathrm{T}}$ data are available, the parameter set of a GMM, $\Gamma = \{\omega_m, \boldsymbol{\mu}_m, \boldsymbol{\sigma}_m \mid m=1,\ldots,M\}$, can be obtained offline by the maximum likelihood estimation technique. This paper uses an off-the-shelf solver, *gmdistribution.fit*, in MATLAB to estimate parameters. Standard guidelines for the GMM and the parameter estimation are available in [33-35].

### C. From wind power to its ramping

Similarly to the net load ramping definition in Eq. (4), the actual wind power ramping $\Delta X$ and the corresponding ramping forecasts $\Delta Y$ are defined as follows:

$$\begin{bmatrix} \Delta \boldsymbol{X} \\ \Delta \boldsymbol{Y} \end{bmatrix} = \Upsilon \begin{bmatrix} \boldsymbol{X} \\ \boldsymbol{Y} \end{bmatrix} \quad (34)$$

where

$$\Upsilon = \begin{bmatrix} \Psi & \\ & \Psi \end{bmatrix} \quad (35)$$

$$\Psi = \begin{bmatrix} -1 & 1 & & & \\ & -1 & 1 & & \\ & & \cdots & & \\ & & & -1 & 1 \end{bmatrix} \quad (36)$$

According to Appendix B, if $[\boldsymbol{X}^{\mathrm{T}} \ \boldsymbol{Y}^{\mathrm{T}}]^{\mathrm{T}}$ is a GMM and $\Upsilon$ is a linear transformation, the distribution of $[\Delta\boldsymbol{X}^{\mathrm{T}} \ \Delta\boldsymbol{Y}^{\mathrm{T}}]^{\mathrm{T}}$ can be computed as follows:

$$f_{\Delta X \Delta Y}\left(\Delta x, \Delta y\right) = \sum_{m=1}^{\mathrm{M}} \omega_m N_m\left(\Delta x, \Delta y \ ; \ \Upsilon\boldsymbol{\mu}_m, \Upsilon\boldsymbol{\sigma}_m \Upsilon^{\mathrm{T}}\right) \quad (37)$$

The distribution of $[\Delta\boldsymbol{X}^{\mathrm{T}} \ \Delta\boldsymbol{Y}^{\mathrm{T}}]^{\mathrm{T}}$ has a GMM form.

### D. Conditional distribution of wind power ramping

For clarity, let $\Upsilon\boldsymbol{\mu}_m$, and $\Upsilon\boldsymbol{\sigma}_m\Upsilon^{\mathrm{T}}$ be reshaped as follows:

$$\Upsilon\boldsymbol{\mu}_m = \begin{bmatrix} \boldsymbol{\mu}_m^{\Delta x} \\ \boldsymbol{\mu}_m^{\Delta y} \end{bmatrix}, \Upsilon\boldsymbol{\sigma}_m\Upsilon^{\mathrm{T}} = \begin{bmatrix} \boldsymbol{\sigma}_m^{\Delta xx} & \boldsymbol{\sigma}_m^{\Delta xy} \\ \boldsymbol{\sigma}_m^{\Delta yx} & \boldsymbol{\sigma}_m^{\Delta yy} \end{bmatrix} \quad (38)$$

If $[\Delta\boldsymbol{X}^{\mathrm{T}} \ \Delta\boldsymbol{Y}^{\mathrm{T}}]^{\mathrm{T}}$ is a GMM, which it is, then the conditional distribution of $\Delta\boldsymbol{X}$ with respect to $\Delta\boldsymbol{Y} = \Delta\boldsymbol{y}$ can be computed shown below [36]. This is also a GMM:

$$f_{\Delta X|\Delta Y}\left(\Delta x \mid \Delta y\right) = \sum_{l=1}^{\mathrm{M}} \omega_l' N_l\left(\Delta x \mid \Delta y \ ; \ \boldsymbol{\mu}_l^{\Delta x \bullet y}, \boldsymbol{\sigma}_l^{\Delta x \bullet y}\right) \quad (39)$$

$$\omega_l' = \omega_l \frac{N_l\left(\Delta y \ ; \ \boldsymbol{\mu}_l^{\Delta y}, \boldsymbol{\sigma}_l^{\Delta y}\right)}{\sum_m \omega_m N_m\left(\Delta y \ ; \ \boldsymbol{\mu}_m^{\Delta y}, \boldsymbol{\sigma}_m^{\Delta y}\right)} \quad (40)$$

where

$$\boldsymbol{\mu}_l^{\Delta x \bullet y} = \boldsymbol{\mu}_l^{\Delta x} + \boldsymbol{\sigma}_l^{\Delta xy}\left(\boldsymbol{\sigma}_l^{\Delta yy}\right)^{-1}\left(\Delta y - \boldsymbol{\mu}_l^{\Delta y}\right) \quad (41)$$

$$\boldsymbol{\sigma}_l^{\Delta x \bullet y} = \boldsymbol{\sigma}_l^{\Delta xx} - \boldsymbol{\sigma}_l^{\Delta xy}\left(\boldsymbol{\sigma}_l^{\Delta yy}\right)^{-1}\boldsymbol{\sigma}_l^{\Delta yx} \quad (42)$$

Denote the net load ramping by $\boldsymbol{Z}$. According to the net load ramping definitions in Eqs. (1)−(4),

$$\boldsymbol{Z} = -\Delta\boldsymbol{X} + \boldsymbol{H} \quad (43)$$

where the $i$th element of $\boldsymbol{H}$ represents the deterministic load ramping, i.e.,

$$\boldsymbol{H}_i = \sum_{d=1}^{D} p_{d,i+1} - \sum_{d=1}^{D} p_{d,i} \quad (44)$$

Once the conditional distribution of wind power ramping is obtained, a conditional distribution of the net load ramping can be computed via Eq. (45), which is a GMM:

$$f_{Z|\Delta Y}\left(z|\Delta y\right) = \sum_{l=1}^{\mathrm{M}} \omega_l' N_l\left(z ; -\boldsymbol{\mu}_l^{\Delta x \bullet y} + \boldsymbol{H}, \boldsymbol{\sigma}_l^{\Delta x \bullet y}\right) \quad (45)$$

**Remark 1**: Random variables $[\boldsymbol{X}^{\mathrm{T}} \ \boldsymbol{Y}^{\mathrm{T}}]^{\mathrm{T}}$, $[\Delta\boldsymbol{X}^{\mathrm{T}} \ \Delta\boldsymbol{Y}^{\mathrm{T}}]^{\mathrm{T}}$, $\Delta\boldsymbol{X}/\Delta\boldsymbol{Y}$, and $\boldsymbol{Z}/\Delta\boldsymbol{Y}$ are GMMs.

**Remark 2**: If the net load ramping is defined not by Eq. (4), but by Eq. (20), the matrix in Eq. (36) should be modified accordingly. Equations (37)−(45) are still applicable.

**Remark 3**: The proposed modeling method considers the aggregated wind power and its ramping. When the chance-constrained zonal reserve requirements and transmission limits are considered, conditional distributions for multiple wind farms are needed. The proposed modeling method should be modified. A detailed discussion is provided in Appendix C.

### E. Comparison of the GMM to prior literature

(1) Many methods, e.g., Copula and Beta, are able to model distributions of wind power [37-39]. Because they cannot compute a linear transformation of a random vector, they are unable to model wind power ramping in an analytical way. In contrast, this paper manages to derive a distribution of ramping from that of wind power because the GMM is used. The GMM has a property called "*linear invariance*" which is shown in Appendix B. On the basis of this property, the ramping distribution is obtained conveniently.

(2) The proposed conditional distribution of wind power ramping depends on forecast values. Therefore, it can be updated dynamically along with the latest forecast information.

(3) The conditional distribution of the net load ramping is a GMM, which benefits the solution of the risk-adjustable FRC allocation problem. This point is detailed in the next section.

## IV. SOLUTION METHODOLOGY

### A. Tractable forms of adjustable chance constraints

With careful derivations, the adjustable chance constraint of Eq. (27) is converted into an equivalent form:

$$\boldsymbol{C}_k^{\mathrm{T}}\boldsymbol{u} + d_k \geq \mathrm{icdf}_{\boldsymbol{D}_k^{\mathrm{T}}\boldsymbol{Z}}\left(1-\alpha_k\right) \quad \forall k \quad (46)$$

Note that $\mathrm{icdf}(\alpha_k)$ is a nonlinear function of $\alpha_k$, which makes the original problem an intractable nonlinear optimization. Considering that $\mathrm{icdf}(\cdot)$ is a single-valued function, this paper uses $\mathrm{icdf}(1-\alpha_k)$ as a new decision variable to substitute for $\alpha_k$, i.e.,

$$\beta_k = \mathrm{icdf}_{\boldsymbol{D}_k^{\mathrm{T}}\boldsymbol{Z}}\left(1-\alpha_k\right) \quad \forall k \quad (47)$$

Thereafter, the adjustable chance constraint is converted into a linear inequality:

$$\boldsymbol{C}_k^{\mathrm{T}}\boldsymbol{u} + d_k - \beta_k \geq 0 \quad (48)$$

Note that there is no need to compute $\alpha_k$ during an iterative solution of the problem. When the iterative solution terminates, one can use the optimal $\beta_k$ and Eq. (47) to find the optimal confidence level $\alpha_k$. A prior task is computing the distribution of $\boldsymbol{D}_k^{\mathrm{T}}\boldsymbol{Z}$. Note that $\boldsymbol{Z}$ is modeled by a GMM, shown in Eq. (45). $\boldsymbol{D}_k$ represents a linear transformation. According to Appendix B, $\boldsymbol{D}_k^{\mathrm{T}}\boldsymbol{Z}$ is a GMM, whose distribution can be computed. Therefore, the optimal confidence level $\alpha_k$ can be obtained.

### B. Computation of potential losses

To compute the integral term in the objective function, *Proposition 1* is proposed (a proof is provided in Appendix D).

**Proposition 1**: if the random vector $\boldsymbol{Z}$ is a GMM, then the integral term in the objective function is computed as follows:

$$\int_{\beta_k}^{\max}\left(v - \beta_k\right)\mathrm{pdf}_{\boldsymbol{D}_k^{\mathrm{T}}\boldsymbol{Z}}\left(v\right)dv =$$
$$= \sum_{m=1}^{\mathrm{M}} \omega_m \left\{ \begin{matrix} \sigma_m^2\left[N_m\left(\beta_k\right) - N_m\left(\max\right)\right] \\ +\mu_m\left[\Phi_m\left(\max\right) - \Phi_m\left(\beta_k\right)\right] \end{matrix} \right\} \quad (49)$$
$$- \beta_k\left[1 - \mathrm{cdf}_{\boldsymbol{D}_k^{\mathrm{T}}\boldsymbol{Z}}\left(\beta_k\right)\right]$$

where $\mu_m, \sigma_m, \Phi_m(\cdot), N_m(\cdot)$ are explained in Appendix D.

As $\boldsymbol{D}_k^{\mathrm{T}}\boldsymbol{Z}$ is a GMM, the $\mathrm{cdf}(\cdot)$ in Eq. (49) can be computed. During an iterative solution, one needs to know the first order derivatives of the objective function. On the basis of **Proposition 1**, the first order derivatives of the integral terms with respect to $\beta_k$ are computed as follows:

$$\frac{\partial}{\partial \beta_k} \int_{\beta_k}^{\max} (v - \beta_k) \, \text{pdf}_{\boldsymbol{D}_k^{\mathrm{T}} \boldsymbol{Z}}(v) \, dv = \text{cdf}_{\boldsymbol{D}_k^{\mathrm{T}} \boldsymbol{Z}}(\beta_k) - 1 \quad (50)$$

**Remark 4**: To derive **Proposition 1**, a very important property of the integral of a Gaussian function is used. That is,

$$\int v N(v) \, dv = -N(v) + C \quad (51)$$

where $N(\cdot)$ is a standard Gaussian distribution and $C$ is a constant.

This paper extends the property of a Gaussian function to a GMM. Many distributions, e.g., Beta and Cauchy, do not have the property depicted in Eq. (51). Therefore, they cannot produce analytical formulae for the integral terms in the objective function. This is one advantage of the GMM.

Many methods [37-39] adopt the Copula method to model the random variable $\boldsymbol{Z}$. As entries of $\boldsymbol{Z}$ are usually correlated, the "convolution technique" does not apply. Consequently, Copula-based methods cannot compute distributions of $\boldsymbol{D}_k^{\mathrm{T}} \boldsymbol{Z}$. In this paper, because $\boldsymbol{Z}$ is a GMM, it is easy to compute the distributions of $\boldsymbol{D}_k^{\mathrm{T}} \boldsymbol{Z}$. This successful attempt benefits from the "*linear invariance*" property of the GMM. Even if the entries of $\boldsymbol{Z}$ are correlated, the "*linear invariance*" is applicable.

In some studies [40], [41], the numerical integral technique or piecewise linear functions are used to compute integral terms of the objective function. In order to achieve high accuracy, integral terms are truncated into many segments. As a result, these two methods are not computationally efficient. In contrast, the method proposed here provides an analytical solution. In this regard, the method is more efficient.

### C.  Comparison to relevant research

(1) The GMM has been used in [33], [34], and [42] to model uncertainties. This paper differs from these studies in two respects. First, in [34] and [42], a GMM is used to represent a univariate PDF of wind power. In contrast, this paper models a joint PDF of multiple random variables. Second, the method of modeling conditional wind power ramping is not reported in [33], [34], or [42].

(2) In [40], Wang et al. combine chance-constrained programming and goal programming to optimize a risk-adjustable UC problem. As the authors of [40] adopt the Gaussian assumption of wind power and the piecewise linearization technique to compute the integral terms, this paper is significantly different from [40].

(3) Compared with the present authors' previous work [43], this paper has three improvements. First, this paper not only models wind power, but also characterizes conditional wind power ramping. Second, by using a fixed confidence level, [43] suffers from the problems that are discussed in Section II-B above. In this paper, the confidence level is adjustable. As a result, the proposed method can automatically achieve a trade-off between economy and risks. Third, the analytical method to compute the potential losses, i.e., the integral terms, is not reported in [43]. These three improvements constitute the major contributions of this paper.

## V.  CASE STUDY

### A.  Data information

The historical wind power data used in this paper is from the "*eastern wind integration data set*" of the National Renewable Energy Laboratory (NREL) [44]. The data records consist of hourly actual wind power outputs and their forecast values. The forecast values are produced using the Weather Research and Forecasting model [45]. There are three forecasting lead time horizons: next-day, 6-hour, and 4-hour. We use the 4-hour ahead data because the proposed method is used for an ED. The data of $2004-2005$ are used as a training set, while the data of 2006 are used as a test set. For demonstration purposes, the number of periods, $I$, is 4. This is a typical time scale for an ED, in which FRC reserves are scheduled.

### B.  Results of modeling wind power ramping

#### 1)  Comparison of a GMM with other distributions

The GMM, with 15 components, is used to model a joint distribution of actual wind power and the forecasts $[\boldsymbol{X}^{\mathrm{T}} \, \boldsymbol{Y}^{\mathrm{T}}]^{\mathrm{T}}$. Then, conditional distributions of wind power ramping $\Delta \boldsymbol{X} / \Delta \boldsymbol{Y}$ are constructed. Empirical distributions are used as benchmarks to evaluate the effectiveness of the GMM-based conditional distributions. The proposed method is compared with three widely used probabilistic models. They are the Gaussian distribution, Beta distribution, and t-Location distribution.

The empirical distributions are obtained as follows:

(a) The historical data pairs of actual wind power and forecasts $[\boldsymbol{X}^{\mathrm{T}} \, \boldsymbol{Y}^{\mathrm{T}}]^{\mathrm{T}}$ are transformed to wind power ramping and ramping forecasts $[\Delta \boldsymbol{X}^{\mathrm{T}} \, \Delta \boldsymbol{Y}^{\mathrm{T}}]^{\mathrm{T}}$.

(b) The data pairs of $[\Delta \boldsymbol{X}^{\mathrm{T}} \, \Delta \boldsymbol{Y}^{\mathrm{T}}]^{\mathrm{T}}$ are divided into several bins on the basis of ramping forecasts. A bin consists of a central value $\Delta y^*$ and a width $wd$. If the ramping forecast value of a data pair is between $\Delta y^*$-$wd$ and $\Delta y^*$+$wd$, this data pair belongs to the bin $[\Delta y^*$-$wd$ , $\Delta y^*$+$wd]$. In this paper, the number of bins is 9. It can be changed to other values if needed. With the NREL data, it is found that the ramping forecast values of most data pairs (99.01%) are between -0.225 and 0.225 p.u. So, the width of each bin is 0.225-(-0.025)/2/9=0.025. For example, the first bin is [-0.225, -0.175], the fifth bin is [-0.025, 0.025], and the last bin is [0.175, 0.225].

(c) Data pairs of the bin $[\Delta y^*$-$wd$ , $\Delta y^*$+$wd]$ are used to extract the empirical distribution of $\Delta \boldsymbol{X}$ conditioned on $\Delta y^*$.

This paper adopts Root Mean Square Error (RMSE) to quantify the fitting performance of different models. A definition of RMSE can be found in [46]. The test results of PDF curves are shown in Fig. 3 and Fig. 4. From Fig. 3, it can be seen that the GMMs match well with empirical distributions. According to Fig. 4, the GMM method has the lowest RMSEs, indicating that the GMM can better represent conditional wind power ramping than the other three models.

Similar tests were also conducted on CDF curves. The RMSEs of CDFs are shown in Fig. 5. The CDF test results coincide with the PDF results, indicating that the GMM outperforms the other three distributions.

From Fig. 3, it can be observed that conditional distributions of wind power ramping are quite different under different forecasts. When the forecasts of wind power ramping increase, the left half-planes of the conditional PDF curves shrink, while the right ones expand.

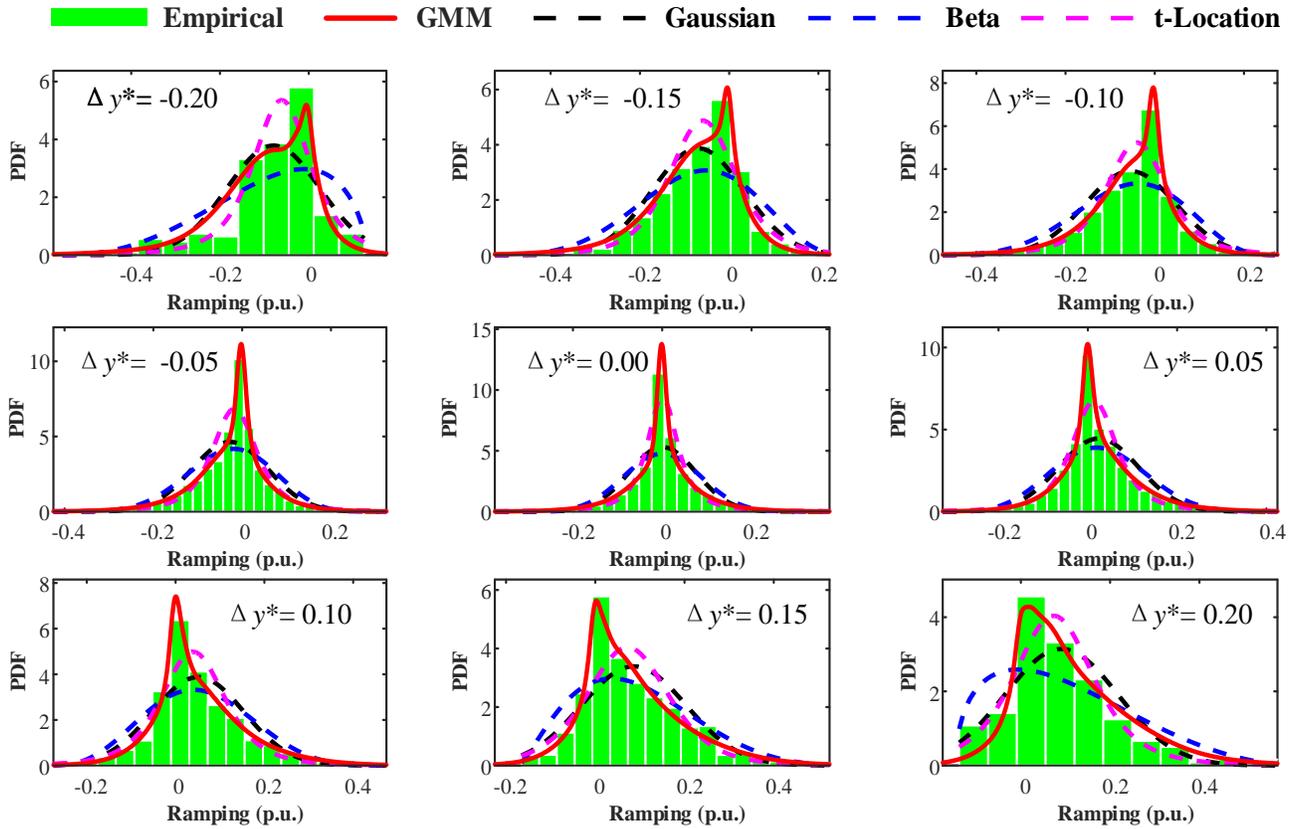

Fig. 3 Conditional distributions of wind power ramping

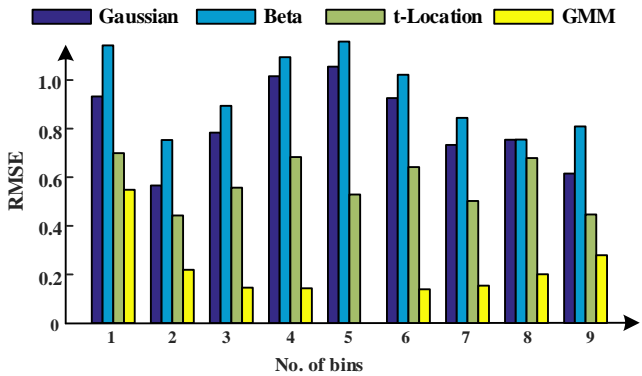

Fig. 4 RMSEs of PDFs in the 9 bins.

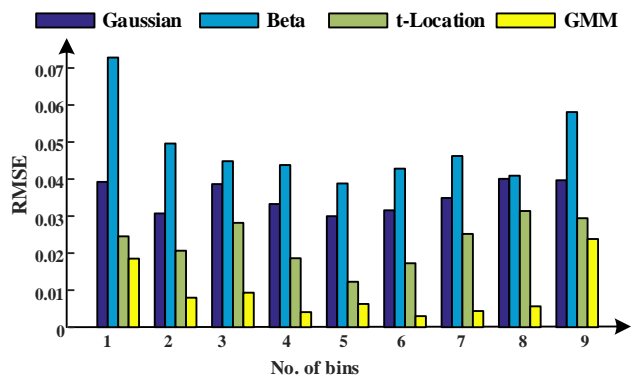

Fig. 5 RMSEs of CDFs in the 9 bins.

### 2) Sensitivity analysis on GMM component size

Usually, a GMM with a small component size may not fairly represent uncertainties. Meanwhile, a large size may reflect overfitting. To determine the GMM component size, this paper computes two indices of a GMM with different component sizes. They are the likelihood function value [47] and average RMSE of PDFs within the 9 bins. The test results are shown in Fig. 6. It can be seen that with 15 components, there is no need to increase the component size, as there is no significant change in the likelihood function value or the average RMSE. That is to say, the fitting performance of the GMM cannot be drastically improved if the component size continues to increase. Therefore, a GMM with 15 components is an appropriate representation for the data series.

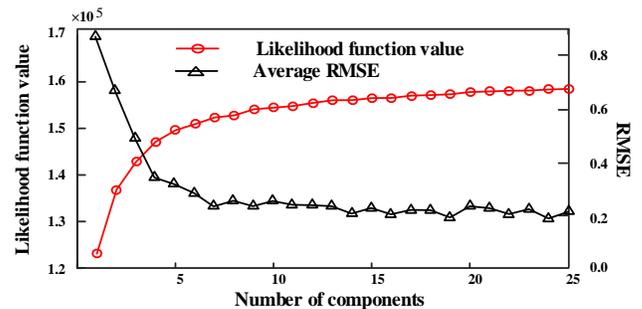

Fig. 6 Sensitivity analysis of component size.

### 3) Test results with BPA data

In addition to the NREL data, which are simulation data, the proposed method was tested with real data from the Bonneville Power Administration (BPA). The dates of the BPA data range from January 1, 2016 to December 30, 2016. The test results are provided in Appendix E. It can be seen that the GMM still achieves a satisfactory modeling performance.

### C. Results of optimal allocation of FRC

The proposed method is tested on a modified IEEE 118-bus system, parameters of which are provided in [48]. There are ten wind farms, each with a capacity of 100 MW. The ten sites are chosen from NREL data series. Their IDs are 3978, 4094, 4208, 4209, 4241, 4429, 4468, 4605, 4703, and 4734. The ten wind farms are geographically close together in Illinois, near longitude 89 West, latitude 41 North. The wind penetration with respect to the base load is 27% (=1000/3668). Penalty coefficients of wind spillage/load shedding ($5/MW) are 5 times higher than FRC cost coefficients ($1/MW).

In the following, a one-month simulation test of the proposed adjustable FRC allocation method is conducted. The indices of the overall performance are computed.

#### 1) Comparison with fixed chance constraints

The purposed of this subsection is to demonstrate that the proposed adjustable chance constraints can benefit the reserve allocation strategy. The proposed adjustable method is compared with a fixed chance-constrained optimization model in which the confidence levels $\alpha$ are 5% (in short, it is called *the fixed method*). Both methods use GMM-based conditional distributions of wind power ramping.

As illustrative examples, the test results for 12 hours are shown in Fig. 7 and Fig. 8. From Fig. 7, the proposed method attains optimal confidence levels $\alpha$ ranging from 5% to 20%, and schedules fewer FRC reserves than the fixed method. According to Fig. 8, compared with the fixed method, the adjustable method has higher potential losses over the 12 periods, but always attains a much lower total cost of the FRC reserve cost and potential loss.

TABLE I
COSTS OF THE FIXED* AND ADJUSTABLE METHODS (10³ $)

| Methods | FRC cost | Wind spillage penalty | Load shedding penalty | Total |
|---------|----------|----------------------|----------------------|-------|
| Fixed | 134.9 | 5.238 | 5.441 | 145.5 |
| Proposed | 82.69 | 15.39 | 13.65 | 111.7 |

*The "fixed" method means the confidence level is 5%.

The results of a one-month simulation test are listed in Table I. Compared with the adjustable method, the fixed method has a much higher FRC cost as well as total cost. Such results indicate that the predefined 5% confidence levels are quite conservative. When the confidence levels are determined by the adjustable method, a better FRC allocation solution with a lower total cost is obtained.

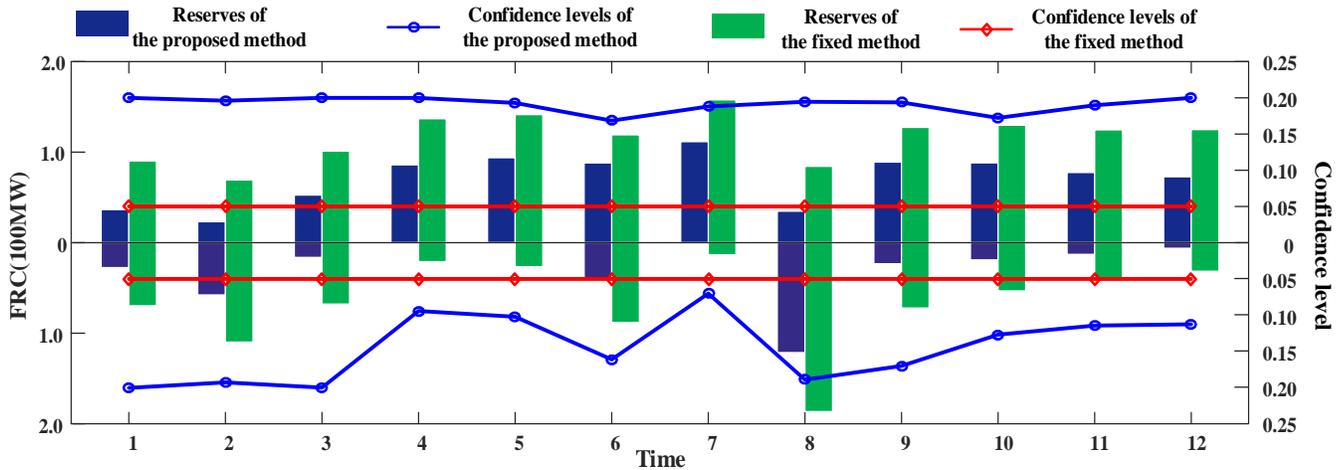

Fig. 7 FRC reserves and confidence levels

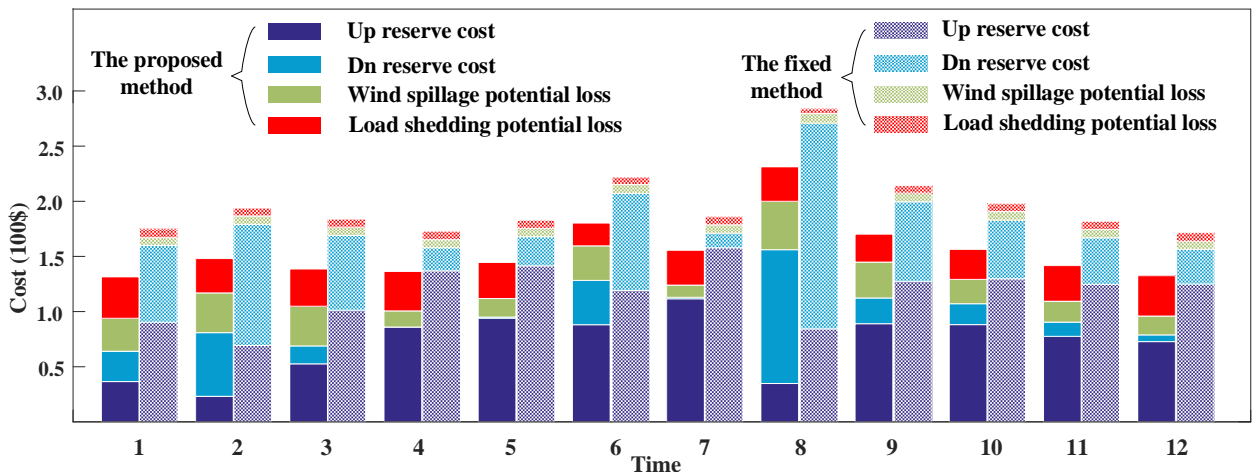

Fig. 8 FRC reserve costs and potential losses

### 2) Comparison with other methods

The proposed method is compared with four methods in the literature, which are briefly described in Table II.

TABLE II
DESCRIPTION OF DIFFERENT METHODS

| Methods | Descriptions |
|---------|--------------|
| *1* | This method is from [49]. FRC reserves are 20% of wind generation installed capacities. |
| *2* | This method is from [19]. Distributions of ramping are Gaussian. Confidence levels $\alpha$ are 5%. |
| *3* | This method combines [19] and [50]. Conditional distributions of ramping are Beta [50]. Confidence levels $\alpha$ are 5%. |
| *4* | This method is from [40]. Uncertainties are modeled by Gaussian distributions. Confidence levels are adjustable. |
| ***Proposed*** | **Conditional GMMs and adjustable confidence levels** |

TABLE III
COSTS OF THE FIVE METHODS ($10^3$\$)

| Methods | FRC reserve cost | Wind spillage penalty | Load shedding penalty | Total |
|---------|------------------|----------------------|----------------------|-------|
| *1* | 287.7 | 1.323 | 0.618 | 289.7 |
| *2* | 172.3 | 7.229 | 5.327 | 184.8 |
| *3* | 128.3 | 23.57 | 16.18 | 168.0 |
| *4* | 102.6 | 19.19 | 14.71 | 136.5 |
| ***Proposed*** | **82.69** | **15.39** | **13.65** | **111.7** |

Cost indices of the overall performance for the five methods are provided in Table III. Four results are obtained:

(1) *Method 1* schedules the most FRC reserves, resulting in the least wind spillage and load shedding. However, it has the highest total cost. That is, *method 1* is not as economical as the other four methods.

(2) Compared with the fixed *method 2*, although the proposed adjustable method has more wind spillage and load shedding, it entails a much fewer FRC reserve cost (52% decrement). As a result, the proposed method has a lower total cost (40% decrement). Compare with *method 3*, the proposed method also attains a lower total cost (33% decrement). From the results, the proposed method outperforms the fixed chance-constrained *methods 2* and *3*.

(3) Compared with the adjustable *method 4* with Gaussian distributions, the proposed method with the GMM-based conditional distributions entails a lower FRC reserve cost (19% decrement), fewer penalties (20% and 7% decrements), and a lower total cost (18% decrement). Such results indicate that the conditional GMM is a more accurate model of ramping uncertainties than the Gaussian distribution.

(4) Among the five methods, the proposed method has the lowest total cost. This demonstrates that the adjustable

approach with the GMM-based conditional distributions can achieve a better overall performance than the other four.

### 3) Sensitivity analysis

#### a) Penalty coefficients

In this test, penalty coefficients for wind spillage and load shedding increase gradually from \$0.5/MW to \$10/MW. FRC reserve cost coefficients remain at \$1/MW. With the different penalty coefficients, the optimal confidence levels and reserves are computed. The test results are shown in Fig. 9. It can be seen that when the penalty coefficients increase, the FRC reserves increase while the confidence levels $\alpha_k$ decrease from 0.45 to 0.08. As a result, there is less wind spillage and load shedding. This test shows that the proposed method is able to adjust itself to avoid high penalties.

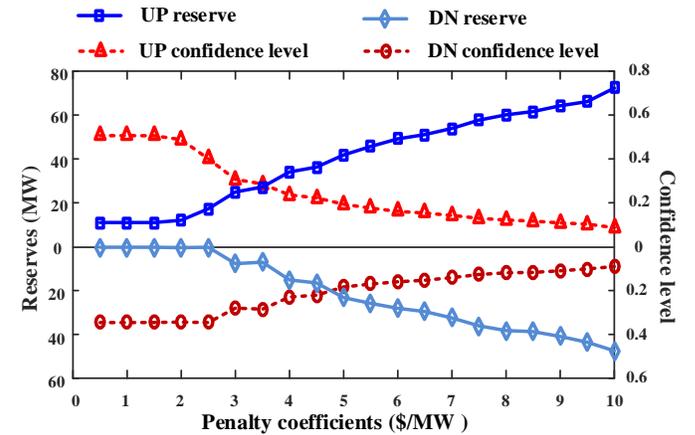

Fig. 9 Optimal confidence levels with different penalty coefficients

#### b) Component size

This paper also computes the objective function values with different GMM component sizes. The test results are shown in Fig. 10. When the component number is over 15, the objective function values do not change significantly. Combining the sensitivity analysis results in Fig. 6 and Fig. 10, this paper suggests that a GMM with 15 components is accurate enough for the uncertainty modeling, and more components will not benefit the optimal solution.

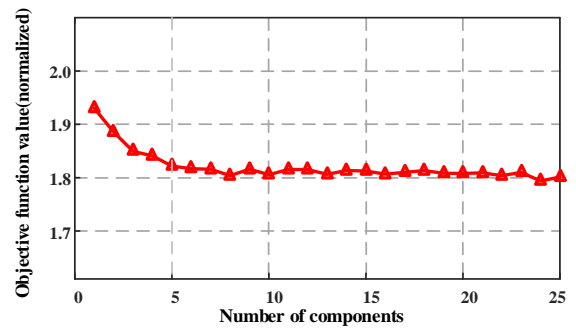

Fig. 10 Sensitivity analysis of the objective function value

All tests are implemented on a PC with a Core-i5 2.39-GHz processor and 8 GB RAM. The coding environment is MATLAB 2013a. The solution of an adjustable FRC allocation problem costs 7.87 seconds.

## VI. CONCLUSION

This paper addresses two important issues: modeling conditional wind power ramping and allocating FRC by an adjustable chance-constrained approach. With the proposed method, it is convenient to construct conditional distributions of wind power ramping under different forecast values. The adjustable chance-constrained approach is able to determine optimal confidence levels of chance constraints, improving the overall performance of an ED with FRC requirements.

The proposed method has two limitations:

(1) The proposed method needs sufficient historical data to estimate the parameters of a GMM. However, data are lacking for some newly built wind farms. In this case, the method is not applicable. A potential solution is to utilize nearby datasets and the Bayesian inference technique [51] to estimate parameters.

(2) The proposed method is not implemented in a distributed manner. As the power system decision-making may evolve from a centralized mode to a decentralized mode [52], [53], future work will entail developing a distributed version of the proposed method.

### APPENDIX A
### EXTENSION TO TRANSMISSION LIMITS
#### 1) Adjustable chance-constrained transmission limits

$$\Pr\left(\begin{array}{c} \sum_{g=1}^{G} s_{l,g} P_{g,i} + \sum_{w=1}^{W} s_{l,w} P_{w,i} \\ -\sum_{l=1}^{D} s_{l,d} P_{d,i} \leq F_{\lim} \end{array}\right) \geq 1 - \alpha_{l,i}^{up} \quad \forall i, \forall l \qquad (52)$$

$$\Pr\left(\begin{array}{c} \sum_{g=1}^{G} s_{l,g} P_{g,i} + \sum_{w=1}^{W} s_{l,w} P_{w,i} \\ -\sum_{l=1}^{D} s_{l,d} P_{d,i} \geq -F_{\lim} \end{array}\right) \geq 1 - \alpha_{l,i}^{dn} \quad \forall i, \forall l \qquad (53)$$

#### 2) Potential losses due to transmission congestion

Suppose that the congestion penalty depends on a fixed coefficient [40]. That is, if the line power is larger than the line limit by an amount $\Delta P_l$, the congestion penalty cost is $\lambda_{con}\Delta P_l$.

Define the weighted aggregation of multiple wind power as follows:

$$p_i^{W,l} = \sum_{w=1}^{W} s_{l,w} p_{w,i} \qquad (54)$$

Thereafter, the potential loss due to congestion is

$$\mathbf{E}\left[f_{Con}\right] = \lambda_{con} \sum_{i=1}^{I} \left[\begin{array}{c} \int_{\beta_{l,i}^{up}}^{\max}\left(v - \beta_{l,i}^{up}\right) \mathrm{pdf}_{p_i^{W,l}}\left(v\right) dv \\ + \int_{\beta_{l,i}^{dn}}^{\max}\left(v - \beta_{l,i}^{dn}\right) \mathrm{pdf}_{-p_i^{W,l}}\left(v\right) dv \end{array}\right] \qquad (55)$$

$$\beta_{l,i}^{up} = \mathrm{icdf}_{p_i^{W,l}}\left(1 - \alpha_{l,i}^{up}\right) \quad , \quad \beta_{l,i}^{dn} = \mathrm{icdf}_{-p_i^{W,l}}\left(1 - \alpha_{l,i}^{dn}\right) \qquad (56)$$

where $\mathrm{pdf}_{p_i^{W,l}}$ is the PDF of $p_i^{W,l}$ and $\mathrm{pdf}_{-p_i^{W,l}}$ is the PDF of $-p_i^{W,l}$.

In addition to transmission limits, the power balance equation (14) can be formulated as an adjustable chance constraint. A discussion is provided below.

#### 3) Adjustable chance-constrained generation adequacy requirement

$$\Pr\left(-p_i^W \leq \sum_{g=1}^{G} P_{g,i} - \sum_{l=1}^{D} P_{d,i}\right) \geq 1 - \alpha_i^{Gen} \quad \forall i \qquad (57)$$

#### 4) Potential loss due to insufficient generation

If the generation is not sufficient, the potential loss is defined as the penalty for load shedding. The potential loss due to insufficient generation is:

$$\mathbf{E}\left[f_{Gen}\right] = \sum_{i=1}^{I} \lambda_{shed} \int_{\beta_i^{Gen}}^{\max}\left(v - \beta_i^{Gen}\right) \mathrm{pdf}_{-p_i^W}\left(v\right) dv \qquad (58)$$

$$\beta_i^{Gen} = \mathrm{icdf}_i^{-W}\left(1 - \alpha_i^{Gen}\right) \qquad (59)$$

where $\mathrm{pdf}_{-p_i^W}$ is the PDF of $-p_i^W$.

### APPENDIX B
### LINEAR INVARIANCE OF GMM [43]

Suppose that a random vector $\boldsymbol{X}$ is distributed by a GMM:

$$f_X\left(\boldsymbol{x}\right) = \sum_{m=1}^{M} \omega_m N_m\left(\boldsymbol{x} \, ; \, \boldsymbol{\mu}_m, \boldsymbol{\sigma}_m\right) \qquad (60)$$

The distribution of a linear transformation $\tilde{\boldsymbol{X}} = \boldsymbol{D}\boldsymbol{X}$ is:

$$f_{\tilde{X}}\left(\tilde{\boldsymbol{x}}\right) = \sum_{m=1}^{M} \omega_m N_m\left(\tilde{\boldsymbol{x}} \, ; \, \boldsymbol{D}\boldsymbol{\mu}_m, \boldsymbol{D}\boldsymbol{\sigma}_m \boldsymbol{D}^{\mathrm{T}}\right) \qquad (61)$$

Equations (60) and (61) hold, regardless of whether entries of $\boldsymbol{X}$ are correlated or not.

### APPENDIX C
### CONDITIONAL MODELS FOR MULTIPLE WIND FARMS

When zonal reserves are considered, a conditional model of $\Delta\boldsymbol{X}/\Delta\boldsymbol{Y}$ for multiple wind farms is needed.

Suppose there are $W$ wind farms. Let $\boldsymbol{X}$ denote the power outputs of multiple wind farms, and $\boldsymbol{Y}$ denote the forecast values:

$$\boldsymbol{X} = \left[X_{1,1} \cdots X_{1,I} \cdots X_{w,1} \cdots X_{w,I} \cdots X_{W,1} \cdots X_{W,I}\right]^{\mathrm{T}} \qquad (62)$$

$$\boldsymbol{Y} = \left[Y_{1,1} \cdots Y_{1,I} \cdots Y_{w,1} \cdots Y_{w,I} \cdots Y_{W,1} \cdots Y_{W,I}\right]^{\mathrm{T}} \qquad (63)$$

For the multiple wind farm case, the derivations to obtain a conditional distribution of $\Delta\boldsymbol{X}/\Delta\boldsymbol{Y}$ are similar to Eqs. (29)−(42). The only difference is that the linear transformation matrix $\boldsymbol{\Upsilon}$ should be modified as follows:

$$\boldsymbol{\Upsilon} = \begin{bmatrix} \boldsymbol{\Psi} & & \\ & \cdots & \\ & & \boldsymbol{\Psi} \end{bmatrix} \qquad (64)$$

The total number of $\boldsymbol{\Psi}$ is $2W$.

When transmission limits are considered, wind power $\boldsymbol{X}$ appears in the adjustable constraints of Eqs. (52) and (53). Therefore, a conditional model of $\boldsymbol{X}/\boldsymbol{Y}$ for multiple wind farms is needed.

Note that the joint distribution of $[\boldsymbol{X}^{\mathrm{T}}\,\boldsymbol{Y}^{\mathrm{T}}]^{\mathrm{T}}$ is represented by a GMM shown in Eq. (31). For clarity, let $\boldsymbol{\mu}_m$, and $\boldsymbol{\sigma}_m$ be reshaped as follows:

$$\boldsymbol{\mu}_m = \begin{bmatrix} \boldsymbol{\mu}_m^x \\ \boldsymbol{\mu}_m^y \end{bmatrix}, \boldsymbol{\sigma}_m = \begin{bmatrix} \boldsymbol{\sigma}_m^{xx} & \boldsymbol{\sigma}_m^{xy} \\ \boldsymbol{\sigma}_m^{yx} & \boldsymbol{\sigma}_m^{yy} \end{bmatrix} \qquad (65)$$

The conditional distribution of $\boldsymbol{X}$ with respect to $\boldsymbol{Y}=\boldsymbol{y}$ can be computed as follows:

$$f_{X/Y}\left(\boldsymbol{x} \mid \boldsymbol{y}\right) = \sum_{l=1}^{M} \omega_l' N_l\left(\boldsymbol{x} \mid \boldsymbol{y} \, ; \, \boldsymbol{\mu}_l^{x*y}, \boldsymbol{\sigma}_l^{xx*y}\right) \qquad (66)$$

$$\omega_l' = \omega_l \frac{N_l\left(\boldsymbol{y} \, ; \, \boldsymbol{\mu}_l^y, \boldsymbol{\sigma}_l^y\right)}{\sum_{m=1}^{M} \omega_m N_m\left(\boldsymbol{y} \, ; \, \boldsymbol{\mu}_m^y, \boldsymbol{\sigma}_m^y\right)} \qquad (67)$$

where

$$\boldsymbol{\mu}_l^{x\bullet y} = \boldsymbol{\mu}_l^x + \boldsymbol{\sigma}_l^{xy}\left(\boldsymbol{\sigma}_l^{yy}\right)^{-1}\left(\boldsymbol{y} - \boldsymbol{\mu}_l^y\right) \quad (68)$$

$$\boldsymbol{\sigma}_l^{xx\bullet y} = \boldsymbol{\sigma}_l^{xx} - \boldsymbol{\sigma}_l^{xy}\left(\boldsymbol{\sigma}_l^{yy}\right)^{-1}\boldsymbol{\sigma}_l^{yx} \quad (69)$$

The conditional distribution of $X/Y$ is a GMM.

### APPENDIX D
### PROOF OF PROPOSITION 1

Note that $\boldsymbol{Z}$ is a GMM, and so is $\boldsymbol{D}_k^\top \boldsymbol{Z}$. Denote the distribution of $\boldsymbol{D}_k^\top \boldsymbol{Z}$ by

$$\mathrm{pdf}_{\boldsymbol{D}_k^\top \boldsymbol{Z}}\left(v\right) = \sum_{m=1}^M \omega_m N_m\left(v\,;\mu_m,\sigma_m^2\right) \quad (70)$$

First, the following equation can be obtained:

$$\int_{\beta_k}^{\max} \beta_k \mathrm{pdf}_{\boldsymbol{D}_k^\top \boldsymbol{Z}}\left(v\right)dv = \beta_k\left[1 - \mathrm{cdf}_{\boldsymbol{D}_k^\top \boldsymbol{Z}}\left(\beta_k\right)\right] \quad (71)$$

Then, a careful derivation is presented as follows:

$$
\begin{aligned}
&\int_{\beta_k}^{\max} v\,\mathrm{pdf}_{\boldsymbol{D}_k^\top \boldsymbol{Z}}\left(v\right)dv \\
&= \int_{\beta_k}^{\max} v\left[\sum_{m=1}^M \omega_m N_m\left(v\,;\mu_m,\sigma_m^2\right)\right]dv \\
&= \sum_{m=1}^M \omega_m \frac{1}{\sqrt{\pi}}\int_{\beta_k}^{\max} v e^{-\left(\frac{v-\mu_m}{\sqrt{2}\sigma_m}\right)^2}d\left(\frac{v-\mu_m}{\sqrt{2}\sigma_m}\right) \\
&= \sum_{m=1}^M \omega_m \left[\begin{aligned}&\frac{\sqrt{2}\sigma_m}{2\sqrt{\pi}}\int_{\beta_k}^{\max} e^{-\left(\frac{v-\mu_m}{\sqrt{2}\sigma_m}\right)^2}d\left(\frac{v-\mu_m}{\sqrt{2}\sigma_m}\right)^2 \\ &+ \frac{1}{\sqrt{\pi}}\mu_m\int_{\beta_k}^{\max} e^{-\left(\frac{v-\mu_m}{\sqrt{2}\sigma_m}\right)^2}d\left(\frac{v-\mu_m}{\sqrt{2}\sigma_m}\right)\end{aligned}\right] \\
&= \sum_{m=1}^M \omega_m \left[\begin{aligned}&\sigma_m^2\left[N_m\left(\beta_k\right) - N_m\left(\max\right)\right] \\ &+ \mu_m\left[\Phi_m\left(\max\right) - \Phi_m\left(\beta_k\right)\right]\end{aligned}\right]
\end{aligned}
\quad (72)
$$

where $\Phi_m$ is the CDF of a normal distribution $N_m(\mu_m, \sigma_m^2)$.

Combining Eqs. (71) and (72), the integral term is obtained:

$$
\begin{aligned}
&\int_{\beta_k}^{\max}\left(v - \beta_k\right)\mathrm{pdf}_{\boldsymbol{D}_k^\top \boldsymbol{Z}}\left(v\right)dv = \\
&= \sum_{m=1}^M \omega_m \left\{\begin{aligned}&\sigma_m^2\left[N_m\left(\beta_k\right) - N_m\left(\max\right)\right] \\ &+ \mu_m\left[\Phi_m\left(\max\right) - \Phi_m\left(\beta_k\right)\right]\end{aligned}\right\} \\
&\quad - \beta_k\left[1 - \mathrm{cdf}_{\boldsymbol{D}_k^\top \boldsymbol{Z}}\left(\beta_k\right)\right]
\end{aligned}
\quad (73)
$$

### APPENDIX E
### TEST RESULTS WITH BPA DATA

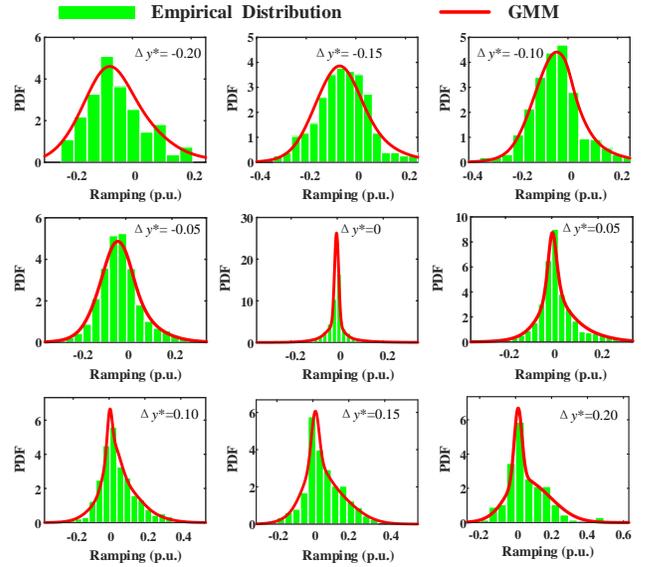

Fig. E-1 Conditional distributions of wind power ramping with BPA data

**Zhiwen Wang** (S'13) received his B.E. degree from the School of Electrical Engineering, Beihang University, Beijing, China, in 2013. He is pursuing his Ph.D. degree in electrical engineering at Tsinghua University, Beijing, China. Currently, he is a visiting scholar at Argonne National Laboratory, Argonne, IL, USA.

**Chen Shen** (M'98–SM'07) received his B.E. and Ph.D. degrees in Electrical Engineering from Tsinghua University, Beijing, China, in 1993 and 1998, respectively. From 1998 to 2001, he was a postdoc in the Department of Electrical Engineering and Computer Science at University of Missouri Rolla, MO, USA. From 2001 to 2002, he was a senior application developer with ISO New England Inc., MA, USA. He has been a Professor in the Department of Electrical Engineering at Tsinghua University since 2009. Currently, he is the Director of Research Center of Cloud Simulation and Intelligent Decision-making at Energy Internet Research Institute, Tsinghua University. He is the author/coauthor of more than 150 technical papers and 1 book, and holds 21 issued patents. His research interests include power system analysis and control, renewable energy generation and smart grids.

**Feng Liu** (M'10) received his B.Sc. and Ph.D. degrees in electrical engineering from Tsinghua University, Beijing, China, in 1999 and 2004, respectively. Dr. Liu is currently an Associate Professor at Tsinghua University. From 2015 to 2016, he was a visiting associate at the California Institute of Technology, CA, USA. His research interests include power system stability analysis, optimal control and robust dispatch, game theory and learning theory and their applications to smart grids. He is the author/coauthor of more than 100 peer-reviewed technical papers and two books, and holds more than 20 issued/pending patents. He is a guest editor of *IEEE Transactions on Energy Conversion*.

**Jianhui Wang** (M'07-SM'12) received his Ph.D. degree in electrical engineering from Illinois Institute of Technology, Chicago, Illinois, USA, in 2007. Presently, he is an Associate Professor with the Department of Electrical Engineering at Southern Methodist University, Dallas, TX, USA. He also holds a joint appointment as Section Lead for Advanced Power Grid Modeling at the Energy Systems Division at Argonne National Laboratory, Argonne, IL, USA. Dr. Wang is the secretary of the IEEE Power & Energy Society (PES) Power System Operations, Planning & Economics Committee. He is an associate editor of Journal of Energy Engineering and an editorial board member of Applied Energy. He has held visiting positions in Europe, Australia and Hong Kong including a VELUX Visiting Professorship at the Technical University of Denmark. Dr. Wang is the Editor-in-Chief of *IEEE Transactions on Smart Grid* and an IEEE PES Distinguished Lecturer. He is also the recipient of the IEEE PES Power System Operation Committee Prize Paper Award for 2015.

**Xiangyu Wu** (S'13) received his B.S. degree from the School of Electrical Engineering, Zhejiang University, Hangzhou, China, in 2012, and his Ph.D. degree in electrical engineering from Tsinghua University, Beijing, China, in 2017. He is currently a postdoc at the School of Electrical Engineering, Beijing Jiaotong University, Beijing, China.